\documentclass[12pt]{article}
\begin{document}
~~~\\
\begin{center}
{\bf Constrained evolution in Hilbert space and requantization\footnote{first printed as preprint FT-380-(1991)/November, at the Institute of Atomic Physics, Bucharest, Romania.} \\[1cm]}
{ M. Grigorescu \\  } 
\end{center}

\noindent
$\underline{~~~~~~~~~~~~~~~~~~~~~~~~~~~~~~~~~~~~~~~~~~~~~~~~~~~~~~~~
~~~~~~~~~~~~~~~~~~~~~~~~~~~~~~~~~~~~~~~~~~~}$ \\[.1cm]
{\bf Abstract} \\
Constrained dynamics on finite-dimensional trial manifolds of quantum state vectors appears in time-dependent variational calculations. This work presents a requantization method, by which  properties of the exact eigenstates can be retrieved from the phase portrait of such approximations. Applications to the coherent states dynamics, small amplitude vibrations and collective rotations, show that this approach extends standard procedures of the quantum many-body theory, such as the method of the projection operators and the random phase approximation.
$\underline{~~~~~~~~~~~~~~~~~~~~~~~~~~~~~~~~~~~~~~~~~~~~~~~~~~~~~~~~
~~~~~~~~~~~~~~~~~~~~~~~~~~~~~~~~~~~~~~~~~~~}$ \\
\newpage \noindent
{\bf Introduction}  \\[.5cm]
Constrained quantum dynamics usually takes into account self-consistent mean-fields in time-dependent variational calculations. Initially it was expressed in the form of the time-dependent Hartree-Fock (TDHF) equations for many-fermion systems. However, the hybrid character of TDHF, including both quantal and classical aspects, leads to difficulties of interpretation \cite{1}. Thus, each TDHF solution is a quantum object, ensuring the optimal description of a time-dependent wave function within the manifold of Slater determinants. Moreover, the small amplitude TDHF vibrations around the critical points represented by solutions of the static Hartree-Fock (HF) equations, correspond to the random phase approximation (RPA). Yet, important properties of the quantum dynamics are lost: the TDHF equations are non-linear, and tunneling effects are suppressed \cite{2}. \\ \indent  
Such classical elements appear not only in TDHF. Any restriction to a manifold of trial states represents more than an approximation, because the result is partly outside the framework of quantum mechanics. Therefore, to get an effective approximation it is necessary to come back in the original Hilbert space by a "requantization" procedure. However, the classical aspects of the mean-field dynamics are useful for the separation of the collective variables \cite{3,4,5,6}.          
\\ \indent
The requantization problem is not yet completely solved, although important partial results have been obtained. The methods of induced representations \cite{7} or geometric quantization \cite{8} can be used to treat globally the peculiar geometry of the collective models, but the constructed Hilbert space is physically independent of the initial one. A different procedure concerns the quantization of the closed orbits, by using a condition of periodicity and "gauge invariance" (GIPQ) \cite{9,10}. This formalism resemble the geometric prequantization \cite{11}, providing integrality constraints of Bohr-Wilson-Sommerfeld type. However, although GIPQ operates within the original Hilbert space, it is still far from complete requantization because the relationship between the periodic trial states selected by such constraints, and the energy eigenstates, remains ambiguous.        
\\ \indent
In this work approximate eigenstates associated to the quantized orbits of time-dependent trial functions are defined by the requantization procedure from \cite{12}. Its relevance was tested on examples, by comparison with other methods and experimental data \cite{th}. The first part contains a short description of the geometric structures related to the constrained quantum dynamics. 
In the second part is presented the formula proposed for the requantization of the periodic orbits. On examples, it is shown that this formula embodies naturally both the RPA and the method of the projection operators. \\[.5cm]
{\bf I. The constrained quantum dynamics} \\[.5cm]
The full quantum dynamics in the Hilbert space ${\cal H}$ is given by the time-dependent Schr\"odinger equation (TDSE)
\begin{equation}
i \hbar \partial_t \vert \psi \rangle = \hat{H}  \vert \psi \rangle
\end{equation}
where $\hat{H}$ is the Hamiltonian operator, defined on a dense domain in ${\cal H}$. This equation can be related to the formalism of classical mechanics by its derivation from the "variational principle" 
$\delta {\cal S}_{[\psi]} =0$, with  ${\cal S}_{[\psi]}=\int dt \langle \psi \vert i \hbar \partial_t - \hat{H}  \vert \psi \rangle$. However, it can also be seen as the flow 
\begin{equation}
 \vert \dot{\psi} \rangle = X_H (\psi)
\end{equation}
of the quantum states $\vert \psi \rangle \in {\cal H}$, under the action of the Hamiltonian field $X_H( \psi ) = (-i / \hbar) \hat{H} \vert \psi \rangle$ defined on ${\cal H}$ by the relation $i_{X_H} \omega^{\cal H}=dh$. Here $h$ denotes the Hamilton function $h(\psi)= \langle \psi \vert \hat{H} \vert \psi \rangle$, $h:{\cal H} \rightarrow R$, and  $\omega^{\cal H} \in \Omega^2({\cal H})$ is the symplectic form (\cite{13}, 4.2):
\begin{equation}
\omega^{\cal H}(X,Y)= 2 \hbar Im \langle X(\psi) \vert Y(\psi) \rangle~~,~~ X(\psi), Y(\psi) \in T_\psi {\cal H}~~.
\end{equation}
In the following ${\cal H}$ is presumed finite dimensional, although part of the results remain valid also in the infinite dimensional case. \\ \indent
The invariance of $h$ to the action on ${\cal H}$ of the group $U(1)$, defined by $\Phi_c \vert \psi \rangle = c \vert \psi \rangle$, $c \in U(1)$, leads to the conservation of the norm $\parallel \vert \psi \rangle \parallel = \sqrt{ \langle \psi \vert \psi \rangle } $. Thus, it is possible to apply the 
Marsden-Weinstein theory to reduce the dynamics from ${\cal H}$ to the projective space ${\bf P}_{\cal H}$ associated to ${\cal H}$ (\cite{14}, 5.5.C). \\ \indent
In terms of $L=\{ \vert \psi \rangle \in {\cal H}, \langle \psi \vert \psi \rangle =1 \}$, and of the projection $\Pi : L \rightarrow {\bf P}_{\cal H} $, 
\begin{equation}
\Pi (\vert \psi \rangle) = [ \vert \psi \rangle ] \equiv \{ e^{i \phi} \vert \psi \rangle, \vert \psi \rangle \in L, \phi \in [0, 2 \pi] \}~~,
\end{equation}
any $\vert Z_t^\gamma \rangle \in L$ having as projection a trajectory $\gamma_t=[\vert Z_t^\gamma \rangle ] \subset {\bf P}_{\cal H}$ of the reduced dynamical system, determines uniquely a solution of the TDSE in ${\cal H}$. This solution has the form 
\begin{equation}
\vert \psi_t \rangle = \Phi_{e^f} \vert Z_t^\gamma \rangle = e^{f} \vert Z_t^\gamma \rangle ~~, \label{5}
\end{equation}
where $f$ is given by
\begin{equation}
\dot{f} \vert Z_t^\gamma \rangle = X_H ( Z_t^\gamma ) - \vert \dot{Z}_t^\gamma \rangle~~.
\end{equation}
Assuming further that $\hat{H}$ is independent of time, then $h=\langle \psi \vert \hat{H} \vert \psi \rangle$ is a constant ${\cal E}$, and $\vert \psi_t \rangle$ from (\ref{5}) becomes:
\begin{equation}
\vert \psi_t \rangle = e^{ - \frac{i}{\hbar} {\cal E} t - \int_0^t d \tau \langle Z_\tau^\gamma \vert
\partial_\tau \vert Z_\tau^\gamma \rangle} \vert Z_t^\gamma \rangle ~~. \label{7}
\end{equation}
This result has a peculiar geometrical interpretation in the prequantization theory. Thus, by the smooth, free and proper action $\Phi : U(1) \times L \rightarrow L $ of the "gauge group" $U(1)$, $L$ is a principal circle bundle over ${\bf P}_{\cal H}$ (\cite{14}, 4.1M). Each normalized state in ${\cal H}$ generates a one-dimensional Hilbert space, and by the natural action of the structure group on this space, $L$ is also the unit section in the associated complex vector bundle \cite{15}. The 1-form $\alpha \in \Omega^1 (L)$, $\alpha_\psi (X)= - i \hbar \langle \psi \vert X \rangle$, $X \in T_\psi L$, ($d \alpha = \omega^{\cal H} \vert_L$), is a connection form on $L$, having as curvature the reduced symplectic form $\tilde{\omega} \in \Omega^2 ({\bf P}_{\cal H})$, $\Pi^* \tilde{\omega} = \omega^{\cal H} \vert_L$. \\ \indent
The covariant derivative \cite{11} of a state $\vert Z \rangle \in L$ along $\gamma \subset {\bf P}_{\cal H}$, $\gamma_t = \Pi ( \vert Z_t \rangle)$, is
\begin{equation}
\frac{ D \vert Z \rangle}{Dt} = \nabla_{\dot{\gamma}} \vert Z \rangle = \frac{i}{\hbar} \alpha ( \vert \dot{Z} \rangle ) \vert Z \rangle = \langle Z \vert \dot{Z} \rangle \vert Z \rangle~~.
\end{equation}
Therefore, in (\ref{7}) the state vector 
\begin{equation}
\vert \tilde{Z}_t^\gamma \rangle = e^{ - \int_0^t d \tau \langle Z_\tau^\gamma \vert
\partial_\tau \vert Z_\tau^\gamma \rangle} \vert Z_t^\gamma \rangle  \label{9}
\end{equation}
is autoparallel along $\gamma$, while  $\vert \psi_t \rangle$ corresponds to the lift of the Hamiltonian current from ${\bf P}_{\cal H}$ to the bundle $L$. \\ \indent
The eigenstates of $\hat{H}$ are critical points of the reduced dynamics on  ${\bf P}_{\cal H}$. In particular, the ground state $\vert \psi_g \rangle$ can be found, up to a phase factor, as the minimum of $h$ on $L$. This property is also used whenever $\vert \psi_g \rangle$ is approximated by the constrained minimum $\vert {\mathsf M} \rangle$ of $h$ on a finite-dimensional "trial" manifold $P \subset L$. In static variational calculations  such as HF or Hartree-Fock-Bogoliubov (HFB), the trial manifolds are orbits of the semisimple Lie groups $SU(n)$ or $SO(2n)$, respectively, generated by unitary action on highest weight vectors for their representation in ${\cal H}$ \cite{16}. Worth noting, the projection $M= \Pi (P)$ for an orbit of this type is a "coherent-state" K\"ahler submanifold of ${\bf P}_{\cal H}$ \cite{17}.   \\ \indent
If $\vert {\mathsf M} \rangle \in P$ is an approximate ground-state, then also the  constrained dynamics on $P$ could be relevant for the description of the quantum system. Let $P$ be a manifold of normalized trial functions $P= \{ \vert Z_{\tilde{x}} \rangle \}$ so that $M= \Pi (P)$ is symplectic. The local coordinates of the point $\Pi (\vert Z_{\tilde{x}} \rangle) \in M$ will be specified by $2N$ real parameters ${\tilde x} \equiv (x^1, x^2,...,x^{2N})$. The Hamilton function on $M$, $h_M : M \rightarrow R$ is the restriction of $h( \psi ) = \langle \psi \vert \hat{H} \vert \psi \rangle$ to $M$. The symplectic form $\omega^M$ of $M$,  $\Pi^* \omega^M = \omega^{\cal H} \vert_P$, defined by the restriction of $\omega^{\cal H}$ to $P$, can be expressed in the form 
\begin{equation}
\omega^M = \sum_{i,j=1}^{2N} \omega_{ij} dx^i \wedge dx^j~~,~~ \omega_{ij}= \hbar Im \langle \partial_i Z \vert \partial_j Z \rangle~~,
\end{equation}
$(\partial_i \equiv \partial / \partial x^i )$. Therefore, the Hamiltonian flow on $M$ is given by the equations 
\begin{equation}
\sum_{j=1}^{2N} 2 \dot{x}^j \omega_{ji} = \frac{ \partial h_M}{\partial x^i} ~~. \label{11}
\end{equation}
The bundle structure $\Pi : L \rightarrow {\bf P}_{\cal H}$ is reproduced at the level of the constrained system only if $M$ is quantizable, so that $L_M \equiv U(1) \cdot P$ is a principal bundle over $M$. In this case the restriction of the 1-form $i \alpha / \hbar \in \Omega^1 (L)$ to $L_M$ yields in the local system of the trial functions $\vert Z \rangle \in P$ the 1-form associated to the connection $\theta_Z \in \Omega^1(M)$, 
\begin{equation}
\theta_Z = \sum_{k=1}^{2N} \langle Z \vert \partial_k \vert Z \rangle dx^k ~~.
\end{equation}
In the prequantization theory, the connection provides the lift the Hamiltonian flow from $M$ to $L_M$. If ${\tilde x}_t$ is a solution of (\ref{11}), its lift $\hat{\rho}_t \vert Z \rangle$ is: 
\begin{equation}
\hat{\rho}_t  \vert Z \rangle = e^{ - \frac{i}{\hbar} {\cal E} t + i \Theta_t} \vert Z_{{\tilde x}_t} \rangle ~~,
\end{equation}
where
$$
{\cal E} = \langle Z_{{\tilde x}_t} \vert \hat{H} \vert Z_{{\tilde x}_t} \rangle~~,~~ \Theta_t = \int_0^t d \tau \langle Z_{{\tilde x}_\tau} \vert i \partial_\tau \vert Z_{{\tilde x}_\tau} \rangle~~. 
$$
The state $\hat{\rho}_t  \vert Z \rangle$ represents a more appropriate approximation to an exact solution of TDSE than $\vert  Z_t \rangle \equiv \vert Z_{{\tilde x}_t} \rangle$. The same result was obtained  
previously \cite{9,10} by using the variational equation 
\begin{equation}
\delta \int {\cal L} ( {\tilde x}, \dot{\tilde x}, r, \dot{r}, \phi, \dot{\phi}) dt =0
\end{equation}
with ${\cal L} = \langle \Psi \vert i \hbar \partial_t - \hat{H} \vert \Psi \rangle $ and $\vert \Psi \rangle = r e^{i \phi} \vert Z_{{\tilde x}_t} \rangle$. Therefore one can say that $\hat{\rho}_t  \vert Z \rangle$ yields the best approximation of a TDSE solution in $L_M$. \\ \indent
It is interesting to remark that $\Theta_t$ coincides with the Berry's phase, derived first for the eigenfunctions of the adiabatic time-dependent Hamiltonians \cite{18,19}, then extended \cite{20} and related to the quantization rules \cite{21}. \\ \indent
In general, the eigenstates of $\hat{H}$ are not elements of $P$, but nevertheless they influence the Hamiltonian flow on $M$. The correspondence between this flow and the energy  eigenstates makes the object of requantization. \\ \indent
The closed orbits on $M$ can be quantized by using the condition of "gauge invariance"\footnote{The invariance up to a constant phase of $\vert \tilde{Z}_t^\gamma \rangle$ in (\ref{9}) when $\vert Z_t^\gamma \rangle$ is replaced by $e^{iS(t)}\vert Z_t^\gamma \rangle$, with $S(t)$ an arbitrary function of time.} and periodicity (GIPQ) \cite{9,10}. This condition was obtained presuming that a periodic trial function resemble the most a time-independent eigenstate. However, beside the argument based on such similarity, GIPQ has a geometrical significance, revealed by the prequantization formalism \cite{th}. Thus, if ${\cal C}$ denotes the set of closed orbits $\gamma_t = \Pi ( \vert Z \rangle^\gamma_t)$ on $M$, then GIPQ approximates the stationary states by the autoparallel vectors: 
\begin{equation}
\vert \tilde{Z}_t \rangle = e^{i \Theta_t} \vert Z^\gamma_t \rangle~~,~~\Theta_t = \int_0^t d \tau \langle Z_\tau^\gamma \vert i  \partial_\tau \vert Z_\tau^\gamma \rangle~~,~~\gamma \in {\cal C} \label{15}
\end{equation}
which are periodic. In the "gauge" of the functions $\vert Z^\gamma \rangle$ having the same period $T_\gamma$ as $\gamma$, denoted $\vert U^\gamma \rangle$, the periodicity condition $\vert \tilde{U}^\gamma_t \rangle=\vert \tilde{U}^\gamma_{t+T_\gamma} \rangle$, $\gamma \in {\cal C}$, leads to the "quantization rule" $\Theta_{T_\gamma}=2 \pi n$, $n \in {\bf Z}$. Stated in terms of the 1-form $\theta_U$, this rule takes the form
\begin{equation}
\frac{i}{2 \pi } \oint_\gamma \theta_U \in {\bf Z}~~. \label{16}
\end{equation}
Unlike the initial periodicity condition, this expression is gauge-dependent. Therefore, it is convenient to use the local relation $\omega^M= - i \hbar d \theta_U$ to express (\ref{16}) in the intrinsic form
\begin{equation}
\frac{1}{h} \int_\sigma \omega^M \in {\bf Z}~~,~~\gamma = \partial \sigma~~.
\end{equation}
Suppose now that the orbits of the Hamiltonian field on the energy surface $P_{\cal E} = h_M^{-1} ({\cal E})$ are closed, and let $\omega_{\cal E}^M$ be the restriction of $\omega^M$ to $P_{\cal E}$. If $\omega_{\cal E}^M$ is regular, then these orbits are also the leafs of the foliation ${\cal F}$ defined by the characteristic distribution of $\omega_{\cal E}^M$ \cite{14}. In this case \cite{22} the condition (\ref{16}) selects the energies ${\cal E}$ for which the quotient spaces $M_{\cal E} \equiv P_{\cal E}/{\cal F}$ are quantizable manifolds.  \\
 \noindent
{\bf II. The requantization of the periodic orbits} \\[.5cm]
If $\vert Z_t \rangle$ is a solution of some time-dependent mean-field equations, the eigenvalues $E_n$ from $\hat{H} \vert \psi_n \rangle= E_n \vert \psi_n \rangle$  can be approximated by  ${\cal E}_n =\langle Z^{\gamma^n} \vert \hat{H} \vert Z^{\gamma^n} \rangle $ for the orbits $\gamma^n \in {\cal C}$ selected by  (\ref{16}). However, if $\vert \tilde{Z}^{\gamma^n} \rangle$ is taken as an approximation for $\vert \psi_n \rangle$, then its time-dependence, though periodic, leads to ambiguities in the evaluation of the transition amplitudes. To avoid such difficulties, in the following will be discussed the possibility of approximating $\vert \psi_n \rangle$ by the time-average\footnote{Knowing from \cite{10} that in some cases the time-averaged overlap between $\vert \tilde{Z}^{\gamma} \rangle$ and $\vert \psi_n \rangle$ attains a maximum for $\gamma =\gamma^n$, I used the time-average to estimate the order of magnitude of some matrix elements  in a computer program \cite{12}. However, it turned out that even the first significant digits were unexpectedly close to the experimental values, suggesting that the time average (\ref{18}) has its own relevance. The result was confirmed by an analytic model (M. Grigorescu, Rev. Roum. Phys. {\bf 34} 1147 (1989)). In 1993 I found the reference S. Dro\.zd\.z, M. Ploszajczak and E. Caurier, Ann. Phys. {\bf 171} 108 (1986), in which a similar time-average was proposed.} $\vert \gamma^n \rangle$ of the periodic vector $\vert \tilde{Z}^{\gamma^n} \rangle$, 
\begin{equation}
\vert \gamma^n \rangle = \frac{1}{T_{\gamma^n}} \oint dt \vert \tilde{Z}^{\gamma^n} \rangle \label{18}
 \end{equation} 
\\[.5cm]
{\it Example 1. The exact stationary solutions} \\[.3cm]
Let ${\cal C}=\{\gamma= \Pi(\vert Z^\gamma \rangle), \vert Z^\gamma_t \rangle = e^{-i \hat{H} t/ \hbar} \vert Z^\gamma_0 \rangle, ~t \in {\bf R} \}$ be a regular orbit cylinder in ${\bf P}_{\cal H}$ \cite{14} corresponding to the exact TDSE solutions. In this case each $\vert Z^\gamma \rangle$ is quasiperiodic, and the related periodic gauge function is a "Bloch wave" in time, $\vert U^\gamma \rangle = e^{i \epsilon^\gamma t / \hbar} \vert Z^\gamma_t \rangle$. The quasienergy $\epsilon^\gamma$ is defined up to integer multiples of $\hbar \omega^\gamma = h /T_\gamma$, and is chosen within the first "Brillouin zone" $[ - \hbar \omega^\gamma /2, \hbar \omega^\gamma /2]$. The condition selecting the quantized orbits $\gamma^n$ becomes ${\cal E}_\gamma = \langle Z^\gamma_0 \vert \hat{H} \vert Z^\gamma_0 \rangle = n \hbar \omega^\gamma + \epsilon^\gamma$, $n \in {\bf Z}$, while  
\begin{equation}
\vert \gamma^n \rangle = \frac{1}{T_{\gamma^n}} \oint dt e^{-it( \hat{H}- {\cal E}_{\gamma^n})/ \hbar} \vert Z^{\gamma^n}_0 \rangle \label{19}    
\end{equation}
represents the projection of $\vert Z^{\gamma^n}_0 \rangle$ on the subspace spaned by the eigenstates with energy ${\cal E}_{\gamma^n}$. The result is close to the ergodic mean theorem (\cite{14}, 3.7.24) and can be easily illustrated for the harmonic oscillator, choosing the orbit cylinder associated to the Glauber coherent states. \\[.5cm]
{\it Example 2. The harmonic oscillator} \\[.3cm]
For the one-dimensional harmonic oscillator $\hat{H}= \hbar \omega (\hat{b}^\dagger \hat{b} + 1/2)$ with $\hat{b}= \sqrt{m  \omega / 2 \hbar} ( \hat{x} + i \hat{p} /m \omega )$. Let $P$ be the manifold of the Glauber coherent states
\begin{equation}
P= \{ \vert Z \rangle = e^{z \hat{b}^\dagger - z^* \hat{b}} \vert 0 \rangle = e^{- \vert z \vert^2/2} e^{z \hat{b}^\dagger} \vert 0 \rangle, z \in {\bf C}, \hat{b} \vert 0 \rangle =0  \} ~~.
\end{equation}
The trajectories on $M= \Pi (P)$ are given by $z_t = e^{-i \omega t} z_0$, 
$\langle Z_t \vert i \partial_t \vert Z_t \rangle = \omega \vert z_0 \vert^2$,
and the condition (\ref{16}) selects the values $z^{(n)}_0$, ${\cal E}_n$:
\begin{equation}
\vert z_0^{(n)} \vert^2 = n \in {\bf N} ~~,~~ {\cal E}_n= \hbar \omega (\vert z_0^{(n)} \vert^2 +\frac{1}{2}) = \hbar \omega (n +\frac{1}{2}) ~~.
\end{equation}
Because $z_t^{(n)} = e^{-i \omega t} z_0^{(n)}$, the period $T$ of $\gamma^n$ is independent of $n$. Thus, the time-average of $\vert \tilde{Z}_t^{\gamma^n} \rangle = \exp (i n \omega t + z_t^{(n)} \hat{b}^\dagger - z^{(n)*}_t \hat{b}) \vert 0 \rangle$ expressed by
\begin{equation}
\vert \gamma^n \rangle = \frac{1}{T} \oint dt \vert \tilde{Z}_t^{\gamma^n} \rangle = 
\frac{e^{-\frac{n}{2}}}{T} \oint dt e^{i n \omega t} e^{z^{(n)}_t \hat{b}^\dagger} \vert 0 \rangle
\end{equation}
yields, up to a normalization factor, the eigenstate $\vert \psi_n \rangle = (n!)^{-1/2} (b^\dagger)^n \vert 0 \rangle$. Worth noting, the "coherence" of the Glauber states makes the lift $\hat{\rho}_t \vert Z \rangle$ an exact solution of TDSE. \\[.3cm]
{\it Example 3. The Random Phase Approximation} \\[.3cm]
Let $G$ be a semisimple Lie group with algebra ${\emph g} \equiv T_e G$, and $M= \Pi (P)$ the K\"ahler submanifold of coherent states projected in ${\bf P}_{\cal H}$ from the group orbit $P=\{ \vert Z \rangle = \hat{U}^a \vert {\mathsf M} \rangle, a \in G,  {\mathsf M} \in {\emph g}^* \}$ of the highest weight vector $\vert {\mathsf M} \rangle \in {\cal H}$. In a local frame of $2n$ real parameters $\{ \xi_j, j=1, 2N \}$, (10) yields $2 \omega_{jk}^M = i \hbar \langle {\mathsf M} \vert [ {\cal D}_j \hat{U}, {\cal D}_k \hat{U} \vert {\mathsf M} \rangle $, and (\ref{11}) becomes 
\begin{equation}
i \hbar \sum_{j=1}^{2N} \dot{\xi}_j \langle {\mathsf M} \vert [ {\cal D}_j \hat{U}, {\cal D}_k \hat{U} \vert {\mathsf M} \rangle = \langle {\mathsf M} \vert [ \hat{H}_U, {\cal D}_k \hat{U} \vert {\mathsf M} \rangle ~~. \label{23}
\end{equation}
Here ${\cal D}_k \hat{U} \equiv \hat{U}^{-1} \partial_k \hat{U}$ and $\hat{H}_U \equiv \hat{U}^{-1} \hat{H} \hat{U}$. In general, to find all periodic solutions of (\ref{23}) is a difficult task, but locally, around the minima of $h_M = \langle {\mathsf M} \vert  \hat{H}_U  \vert {\mathsf M} \rangle$, the periodic orbits exist and can be obtained by integrating its linearized form. \\ \indent
If $\Delta \in {\emph g}^*$ denotes the set of roots for ${\emph g}$, $\Sigma = \{ {\mathsf m} \in \Delta, ({\mathsf m}, {\mathsf M} ) < 0 \}$, and the operators $\{  \hat{E}_{\mathsf m}, \hat{H}_{\mathsf m}, {\mathsf m} \in \Delta \}$, 
$$
\hat{E}_{\mathsf m}^\dagger=\hat{E}_{-{\mathsf m}}~~,~~ \hat{H}_{\mathsf m} \vert {\mathsf M} \rangle = ( {\mathsf m}, {\mathsf M}) \vert {\mathsf M} \rangle ~~,    
$$
represent in ${\cal H}$ the Cartan-Weyl basis for ${\emph g}$, then around a critical point $\hat{U}_0  \vert {\mathsf M} \rangle \in P$ the parameters $\xi_j$ can be chosen as the real and imaginary parts $x_{\mathsf m}$, $y_{\mathsf m}$ of the complex variables $z_{\mathsf m}$, ${\mathsf m} \in \Sigma$ defined by
\begin{equation}
\vert Z \rangle = \hat{U}_z \vert {\mathsf M} \rangle~~,~~ \hat{U}_z = \hat{U}_0 e^{ \sum_{{\mathsf m} \in \Sigma} (z_{\mathsf m} \hat{E}_{\mathsf m} - z_{\mathsf m}^* \hat{E}_{-{\mathsf m}})}~~.
\end{equation}
With these parameters, the components of the form $\omega_{ij}^M$ in $\hat{U}_0 \vert {\mathsf M} \rangle$ are $\omega_{x_{\mathsf m} x_{\mathsf n}}^M = \omega_{y_{\mathsf m} y_{\mathsf n}}^M =0$, and
\begin{equation}
\omega_{x_{\mathsf m} y_{\mathsf n}}^M = - \hbar \langle {\mathsf M} \vert [ {\cal D}_{z_{\mathsf m}} \hat{U}, {\cal D}_{ z_{- {\mathsf n}}} \hat{U} \vert {\mathsf M} \rangle = - \hbar \delta_{{\mathsf m} {\mathsf n}}
\langle {\mathsf M} \vert [ \hat{E}_{\mathsf m}, \hat{E}_{- {\mathsf n}} ] \vert {\mathsf M} \rangle   ~~.
\end{equation}
Retaining only the terms linear in $z$, the equations of motion  (\ref{23}) become 
\begin{equation}
i \hbar \dot{z}_{\mathsf m} \langle {\mathsf M} \vert [ \hat{E}_{\mathsf m}, \hat{E}_{-{\mathsf m}} \vert {\mathsf M} \rangle =
\langle {\mathsf M} \vert [ \hat{H}_{U_0} , \hat{E}_{-{\mathsf m}} \vert {\mathsf M} \rangle - \label{26}
\end{equation}
$$
\sum_{{\mathsf n} \in \Sigma} \langle {\mathsf M} \vert 
[[(z_{\mathsf n} \hat{E}_{\mathsf n} - z_{\mathsf n}^* \hat{E}_{-{\mathsf n}}), \hat{H}_{U_0}], \hat{E}_{-{\mathsf m}}]
\vert {\mathsf M} \rangle ~~.
$$
The state $\vert g \rangle = \hat{U}_0 \vert {\mathsf M} \rangle$ is a critical point if the constant term in (\ref{26}) vanishes, 
\begin{equation}
\langle {\mathsf M} \vert [ \hat{H}_{U_0} , \hat{E}_{-{\mathsf m}} ] \vert {\mathsf M} \rangle =0~~,~~ \forall {\mathsf m} \in \Sigma~~, 
\end{equation}
which is the static equation used to find $\hat{U}_0$. However, further will be taken $\hat{U}_0 \equiv \hat{I}$, presuming that if the minimum exists, then it represents both the ground state and  $ \vert {\mathsf M} \rangle \in P$. Moreover, $\vert g \rangle$ is supposed to be invariant to the symmetry group $G_S$ of $\hat{H}$. Otherwise, the action of the identity component of $G_S$ on $\vert g \rangle$ generates a whole critical manifold for $h_M$, related to the collective modes. Important examples are provided by the anisotropic or superfluid ground states. These are not invariant to the action of the rotation group $SO(3, {\bf R})$, respectively of the "gauge" group $U(1)$ generated by the particle number operator, and the collective behavior shows up in rotational bands. \\ \indent
Let $\vert {\mathsf M} \rangle$ be a $G_S$-invariant minimum, and consider the closed orbits associated to the normal vibration modes. These are specified by the complex amplitudes $\{ X_{\mathsf m}, Y_{\mathsf m}, {\mathsf m} \in \Sigma \}$ and the oscillation period $T$, of the general periodic solution 
\begin{equation}
z^\omega_{\mathsf m} = X_{\mathsf m} e^{- i \omega t}+ Y_{\mathsf m} e^{i \omega t} ~~,~~ {\mathsf m} \in \Sigma,~ \omega = \frac{2 \pi}{T}~~. \label{28}
\end{equation}
Replacing this expression in (\ref{26}), a time-independent system is obtained:
\begin{equation}
\langle {\mathsf M} \vert 
[[ \hat{H}, \hat{B}^\dagger] - \hbar \omega \hat{B}^\dagger, \hat{E}_{\mathsf m}] \vert {\mathsf M} \rangle =0~~,~~
\forall {\mathsf m} \in \Delta,~({\mathsf m},{\mathsf M}) \ne 0~, \label{29}
\end{equation} 
where $\hat{B}^\dagger$ denotes the sum
\begin{equation}
\hat{B}^\dagger= \sum_{{\mathsf m} \in \Sigma} ( X_{\mathsf m} \hat{E}_{\mathsf m}- Y_{\mathsf m}^* \hat{E}_{- {\mathsf m}})~~.
\end{equation}
For (\ref{28}),  $\theta_U=     \langle {\mathsf M} \vert  \hat{U}^{-1} \partial_\tau \hat{U} \vert {\mathsf M} \rangle$ from (\ref{16}) is dominated by the term 
\begin{equation}
\sum_{{\mathsf m} \in \Sigma} \langle {\mathsf M} \vert [ \hat{E}_{\mathsf m}, \hat{E}_{- {\mathsf m}}] \vert {\mathsf M} \rangle Im (\dot{z}_{\mathsf m} z^*_{\mathsf m} ) =
\end{equation}
$$
- \omega \sum_{{\mathsf m} \in \Sigma} \langle {\mathsf M} \vert [ \hat{E}_{\mathsf m}, \hat{E}_{- {\mathsf m}}] \vert {\mathsf M} \rangle ( \vert X_{\mathsf m} \vert^2- \vert Y_{\mathsf m} \vert^2) ~~,
$$
and the quantization condition for the amplitudes $X_{\mathsf m}, Y_{\mathsf m}$  becomes
\begin{equation}
- \sum_{{\mathsf m} \in \Sigma} \langle {\mathsf M} \vert [ \hat{E}_{\mathsf m}, \hat{E}_{- {\mathsf m}}] \vert {\sf M} \rangle ( \vert X_{\mathsf m} \vert^2- \vert Y_{\mathsf m} \vert^2) =n, ~ n=1,2, ... \label{32}
\end{equation}
The result makes sense only if the quantized amplitudes are within the range of the linear approximation. Assuming this to be true for $n=1$, the autoparallel section with period $T$ becomes 
\begin{equation}
\vert \tilde{Z}_t^\omega \rangle = e^{ i \omega t} e^{e^{-i \omega t} \hat{B}^\dagger - e^{i \omega t} \hat{B}} \vert {\mathsf M} \rangle
\end{equation}
and the approximate stationary state given by the time-average is 
\begin{equation}
\vert \omega \rangle = \frac{1}{T} \oint \vert \tilde{Z}_t^\omega \rangle dt \approx \hat{B}^\dagger 
\vert {\mathsf M} \rangle~~.
\end{equation}
In the HF case $G= SU(n)$, and the set of operators associated to the roots ${\mathsf m} \in \Sigma$ contains $A(n-A)$ operators $\{ \hat{c}^\dagger_i \hat{c}_j, A+1 \le i \le n, 1 \le j \le A \}$ expressed in terms of the single-particle creation and annihilation fermion operators $\{ \hat{c}^\dagger_i, \hat{c}_i, i=1,n \}$. In this representation (\ref{29}) becomes the RPA equation in the particle-hole channel, and for the first excited state ($n=1$) the quantization condition (\ref{32}) reduces to the normalization equation for the RPA operators\footnote{This relation between quantization and normalization was obtained using a different method by S. Levit, J. W. Negele and Z. Paltiel, Phys. Rev. C {\bf 21} 1603 (1980), Eq. 3.39.}, $\langle {\mathsf M} \vert [ \hat{B}, \hat{B}^\dagger] \vert {\mathsf M} \rangle =1 $. \\ \indent
Similarly, by choosing $G=SO(2n)$  the RPA equations in the particle-particle channel\footnote{If $\vert {\mathsf M} \rangle$ is the particle vacuum, for ${\mathsf m} \in \Sigma $ there are $n(n-1)$ operators $\{ \hat{c}^\dagger_i \hat{c}_j^\dagger, i,j=1,n \}$. An example is presented in M. Grigorescu and E. Iancu, Z. Phys. A {\bf 337} 139 (1990). } are obtained. \\[.3cm]
{\it Example 4. The requantization of the periodic orbits in action-angle coordinates} \\[.3cm]
Let the dynamical system defined on $M= \Pi (P)$ be completely integrable in the action-angle coordinates $(\tilde{I}, \tilde{\varphi})$, $\tilde{\varphi}=(\varphi^1, \varphi^2, ..., \varphi^N) \in {\bf T}^N$, $\tilde{I} = (I_1, I_2,...,I_N) \in {\bf R}^N$.  
The manifold $P$ will be represented as 
\begin{equation}
P= \{ \vert Z( \tilde{\varphi}, \tilde{I}) \rangle = e^{- \frac{i}{\hbar} \sum_{k=1}^N \varphi^k \hat{J}_k} 
\vert Z (0, \tilde{I}) \rangle \} \label{4p}
\end{equation}
with $[\hat{J}_i, \hat{J}_k ]=0$. If there is no degeneracy the closed orbits correspond to the fundamental cycles $\{ \gamma_k, k=1,N \}$ of $M$, $\dot{\varphi}^j \vert_{\gamma_k} = \delta_{jk} \lambda^k$. The condition (\ref{16}) leads to the quantized expectation values  
\begin{equation}
\langle Z^{\gamma_k} \vert \hat{J}_k \vert Z^{\gamma_k} \rangle =  m \hbar~~,m \in {\bf Z}
\end{equation}
and further to a quantized set of action variables $\tilde{I}$. \\ \indent
The autoparallel vectors defined by (\ref{15}) are 
\begin{equation}
 \vert \tilde{Z}^{\gamma_k} \rangle = e^{im \varphi^k(t)}  e^{- \frac{i}{\hbar} \varphi^k(t) \hat{J}_k}
 \vert Z^{\gamma_k} (0, \tilde{I}) \rangle
\end{equation}
and their average over the orbit $\gamma_k^m$:
\begin{equation}
 \vert \gamma_k^m \rangle =  \frac{1}{2 \pi} \oint 
d \varphi  e^{- \frac{i}{\hbar} \varphi (\hat{J}_k -m \hbar) }
 \vert Z^{\gamma_k^m} (0, \tilde{I}) \rangle
\end{equation}
is the projection of the state $\vert Z^{\gamma_k^m} (0, \tilde{I}) \rangle$ on the subspace spaned by the eigenstates of $\hat{J}_k$ with the eigenvalue $m \hbar$. \\ \indent
In applications, constraining manifolds $P$ of the form (\ref{4p}) appear when the variational equation 
$\delta \langle Z \vert \hat{H} \vert Z \rangle=0$ on a general trial manifold $V$ has a "deformed" (symmetry-breaking) solution $\vert g \rangle$. Let $T^N$ be the maximal torus of the symmetry group $G_S$ of $\hat{H}$.  The ground-state $\vert g \rangle \in V$ is presumed non-invariant to the action of any subgroup of $T^N$, and is used to define the point $\vert Z (0,0) \rangle \equiv \vert g \rangle$ of $P \subset V$. The states $\vert Z (\tilde{\varphi}, \tilde{I}) \rangle$ are related to $\vert Z (0, \tilde{I}) \rangle$ by the action of $T^N$, and therefore to define $P$ is necessary to specify the leafs parameterized by the action variables. Because 
$$
2 \omega_{\varphi^j I_k} = 2 \hbar Im \langle \partial_{\varphi^j} Z (\tilde{\varphi}, \tilde{I}) \vert \partial_{I_k} Z (\tilde{\varphi}, \tilde{I}) \rangle = \partial_{I_k} 
  \langle  Z (\tilde{\varphi}, \tilde{I}) \vert \hat{J}_j \vert Z (\tilde{\varphi}, \tilde{I}) \rangle ~~,
$$
the canonical expression $2 \omega_{\varphi^j I_k} = \delta_{jk}$ is obtained if  
$$
I_k=  \langle  Z (\tilde{\varphi}, \tilde{I}) \vert \hat{J}_k \vert Z (\tilde{\varphi}, \tilde{I}) \rangle =
  \langle  Z (0, \tilde{I}) \vert \hat{J}_j \vert Z (0, \tilde{I}) \rangle ~~.
$$ 
Moreover, because $[\hat{H}, \hat{J}_k]=0$ the Hamiltonian 
$$
h_M = \langle  Z (\tilde{\varphi}, \tilde{I}) \vert \hat{H} \vert Z (\tilde{\varphi}, \tilde{I}) \rangle =
  \langle  Z (0, \tilde{I}) \vert \hat{H} \vert Z (0, \tilde{I}) \rangle 
$$
depends only on $\tilde{I}$. Thus, for a given $\tilde{I}$, the state $\vert Z (0, \tilde{I}) \rangle$ can be taken as the minimum of  $h_M=\langle  Z \vert \hat{H} \vert Z \rangle $ on the submanifold of $V$ selected by the set of $N$ conditions $I_k=  \langle  Z \vert \hat{J}_k \vert Z \rangle$, $k=1,N$. This constrained variational problem reduces to a normal one for the modified Hamiltonian
\begin{equation}
\hat{H}_{\tilde{\lambda}}= \hat{H}- \sum_{k=1}^N \lambda^k \hat{J}_k
\end{equation}
where $ \tilde{\lambda} = \{ \lambda^k, k=1,N \}$ are the Lagrange multipliers. Denoting by $\vert  g_{\tilde{\lambda}} \rangle$ the solution of the variational equation $\delta \langle Z \vert \hat{H}_{\tilde{\lambda}} \vert Z \rangle=0$ in $V$, and by $\lambda^k(\tilde{I})$ the functions of $\tilde{I}$ determined by the system of implicit equations $I_k=  \langle  g_{\tilde{\lambda}} \vert \hat{J}_k \vert g_{\tilde{\lambda}} \rangle$, $k=1,N$, the result is $\vert Z (0, \tilde{I}) \rangle \equiv \vert g_{\tilde{\lambda} (\tilde{I} )} \rangle$. \\[.5cm]
{\bf Discussion and conclusions} \\[.5cm]
The natural restriction of the quantum dynamics to the manifold of normalized states, related to the phase invariance, was used as a model also for the finite-dimensional, quantizable, trial manifolds. \\ \indent
The correspondence between the constrained dynamics and the exact solutions of TDSE was studied for Hamiltonian operators independent of time. It was shown that the closed orbits on the trial manifold can be related to the stationary states by a procedure extending the prequantization formalism and the GIPQ method. This procedure requires:
\begin{itemize}
\item The selection of the "quantizable" closed orbits lying under periodic autoparallel sections.
\item The calculation of approximate stationary states by the time-average of the selected autoparallel sections. 
\end{itemize} 

In applications the choice of the constraining manifolds plays a central role. Particularly important appear to be the exact orbit cylinders and the manifolds associated to the symmetry-breaking ground states. Depending on this choice, the proposed procedure appears as a fine instrument either for the study of the exact quantum dynamics, or of the collective modes. In particular, it shows the common nature of different techniques such as the restoration of broken symmetries by using projection operators, and the random phase approximation. \\ \indent
It is important to remark that for the harmonic oscillator, among all trial manifolds the Glauber coherent states are distinguished because the constrained dynamics coincides with the one of the  classical system. Therefore, in this case the problem of quantization can be stated as the requantization of the constrained dynamics, rather than its correspondence with an isomorphic, but different quantum framework. \\ \indent
These results indicate the consistency and physical relevance of the proposed procedure. However, this should be considered only as a part of a general requantization formalism. A straightforward extension concerns the requantization of the orbits dense on invariant phase-space tori, and will be considered in a subsequent paper.

\end{document}